# Using clocks to determine the entropy of black holes and other space-time singularities.

Dr. Akinbo Ojo
*Standard Science Center,*
*P.O. Box 3501, Surulere, Lagos, Nigeria.*
*Email: taojo@hotmail.com*
Tel. 234-1-4345025

# Abstract

Space-time singularities, viz. Big bang, Big crunch and black holes have been shown to follow from the singularity theorems of General relativity. Whether the entropy at such infinite proper-time objects can be other than zero has also been a longstanding subject of research. Currently the property most commonly chosen to calculate their entropy is a multiple of the surface area of the event horizon and usually gives non-zero entropy values. Though popular, this choice still leaves some substantial questions unanswered hence the motivation for alternative methods for entropy derivation. Here, we use a different property, the proper-time at singularities based on the General relativity predicted behavior of clocks, to derive their entropy. We find, firstly within statistical and thermodynamic principles, secondly when this property is taken into account in the Bekenstein-Hawking formula and thirdly illustrating with a natural analogue, that the entropy of black holes and all other gravitational singularities cannot be other than zero as had been earlier classically speculated.

# Introduction

Calculating what the entropy of black holes and other space-time singularities of General relativity (GR) should be has been a subject of over thirty years controversy and research. It is strongly suspected that a thorough understanding of the subject will throw some collateral light on some other outstanding fundamental questions challenging physics [1]. A number of parameters of gravitational singularities such as the event horizon, surface gravity, thermal atmosphere, mass, charge, thermal equilibrium, etcetera have been used to describe the thermodynamics of black holes in particular [1]. The use of a multiple of the surface area of the event horizon to calculate the entropy of a black hole [2,3] is the currently most favored parameter and it has changed the classical values of zero entropy and absolute zero temperature which had previously been associated with black holes [4,5]. However a number of open issues and paradoxes remain with this choice [1]. Some of these include the absence of clear statistical variables accounting for the entropy and the consequent absence of a logarithmic term necessary to preserve the statistical interpretations of entropy; the choice tells us nothing of the effect that the deformity in space and the time will have on the entropy of singularities, i.e. whether the way clocks run has any thermodynamic effect or not; whether the area law being a theorem in differential geometry has a strong enough link with dynamics or statistics and thus if it will be sensitive enough as a marker for entropy is uncertain; there is also remaining uncertainty whether the area law is enough to specify the entropy for all singularities, i.e. whether initial singularities like the Big bang also have event horizons or not; we do not know whether the Big bang and Big crunch cosmological singularities will have the same event horizon surface area making their entropies of equal value, even though the second law of thermodynamics suggests that the entropy at the Big bang was low; it is uncertain whether or not there can be Hawking radiation to emit the universe from the Big bang in a time reverse of the Big crunch [6,7,8] when there may be no exterior space outside the singularity where particle-antiparticle pairs could be created to make this happen according to Hawking's postulation; we do not know whether the area law can also be applicable in calculating the entropy of those gravitational singularities formed where no matter is involved at all, (e.g. singularities formed by gravitational wave focusing [9,10,11]) and many other remaining questions. These important unresolved problems provide motivation for a deeper understanding of entropy and a search for alternative derivations of black hole entropy from first principles in General relativity and thermodynamics, possibly using what can be universally applicable to all types of space-time singularities. Although this paper may have other wide ranging implications for our understanding of the structure of singularities, their mechanism of formation, cosmology and time symmetry, their entropy will be the main focus of this particular paper.

Entropy cannot be measured directly but must instead be calculated from parameters that can be measured. It is noteworthy that time is a recurring parameter in the characterization of entropy (e.g. in the second law of thermodynamics, 'dynamics' itself implying time evolution of a parameter). This may make time a sensitive marker for entropy. In turn, Einstein's theory of relativity [8,12] gave us novel perspectives on how "clocks" run under certain gravitational situations. Time is therefore common and of essence to both relativity and thermodynamics. Here, we propose the use of proper-time and how clocks run as a parameter to determine the value of the entropy of black holes in particular and other singularities in general.

The inevitability for the occurrence of gravitational or space-time singularities follow from the singularity theorems of Hawking and Penrose [13,14] within the framework of Einstein's theory of gravitation. They can be divided into initial space-time singularities which precede the appearance of space-time and matter (e.g. Big bang) and final singularities which are the end result of disappearance of space-time and matter (e.g. Big crunch, black holes). In other words, singularities represent an absence of space-time [7,13,14]. The definition of a space-time singularity is essentially a dynamic one concerning the geometric behavior of light propagation, light playing a crucial central role in both Special and General relativity. Singularities are fundamentally states or phenomena where space-time is infinitely curved. In order to further remove ambiguity in terminology, a region of infinitely curved space-time is one in which the proper-time taken by light to travel a given distance is infinitely prolonged or dilated [12,13,14]. In other words, nothing, not even light can muster the velocity that can escape such objects. This infinite proper-time property is general to all gravitational singularities whether initial or final, stationary or rotating, cosmological or local, large or small, massive or even massless (as in [9,10,11]) making it a marker with some specificity for singularities. General relativity had taught us that contrary to earlier notions, all clocks do not run at the same rate but depend on the effect of the gravitational field where they are located. The time read by a clock at a given location is the proper-time at that location and it is the behavior of clocks at that location that describes the physics and not the behavior of clocks elsewhere. Slowing of clocks based on the gravity of where they are located is experimentally verified, as light takes a longer time to transit in such areas [15], depending on the strength of the gravitational field there. In singularities, General relativity tells us that not only do clocks run slow, clocks actually come to a stand still!! It is reasonable from the foregoing to suspect that how clocks behave within singularities may have some impact on their thermodynamics and entropy.

There have however been few discussions in the literature of what would happen to thermodynamics when GR is taken into account and when discussed reservations are usually expressed due to some speculated problems. A notable brief discussion can be found in [16]. Despite this avoidance, it is difficult to imagine how the thermodynamics of black holes can be fully and accurately described without input from GR from which they were constructed in the first place. Part of the result of this avoidance is the predominance of efforts at calculating the entropy "residing" on the outside or horizon of the hole and not in the "deep interior" as should probably rightly be the case as suggested by Wald [16].

Some of the reasons given for partially excluding GR is that GR is a field theory and as such it ascribes infinitely many degrees of freedom to the space-time metric/ gravitational field and thus no sensible thermodynamics should be possible. If however space-time possesses granularity at the Planck scale as is strongly suggested will be the case in a correct quantum gravity theory, then the degrees of freedom of the space-time/gravitational field can be finite and sensible thermodynamics that is compatible with GR will be possible. Also although the validity of thermodynamics for ordinary systems is based on the presence of a well-defined notion of time translations and in GR there is an absence of any such rigid time translation structure, thankfully in black holes and other gravitational singularities the time translation structure is rigid and known and according to the singularity theorems of GR this is infinite [13, 14]. It is the contention in this paper that the "peculiar" time translation structure present in black holes, etc will be relevant in understanding and calculating their entropy and should not be ignored. Finally the singularity theorems and the Friedmann-Robertson-Walker space-time models tell us that singularities are present at the end of space-time collapse (e.g. black holes, Big crunch) and at the beginning of evolution of collapsed states to uncollapsed configurations (e.g. Big bang and the hypothetical white holes) and so feared violations of ergodic behavior [16] that prevent evolution of singularities to uncollapsed states may not really exist in classical GR. Consequent to all the above, it can be optimistically speculated that many of the perceived obstacles in fully incorporating GR in our study of the thermodynamics of black holes and other singularities can and indeed must be overcome if we are to have a full and comprehensive picture of their thermodynamics.

Black holes having zero entropy is not new. Before the Bekenstein proposal [2], black holes (final singularities) were taken to likely have zero entropy [4,5], which is identical to that of the Big bang (an initial singularity) as suggested by the second law of thermodynamics. Moreover "extremal" black holes are described and accepted to have zero entropy [1] in spite of their having event horizons. In view of the time-symmetrical nature of "all the known physics" involved (thermodynamics, general relativity, Friedmann-Robertson-Walker space-time models, standard unitary procedures of quantum theory) it will indeed be expected that all singularities, initial or final, will be somewhat identical and have the same entropy as noted by Penrose [7]and others. If it will be necessary for dissimilarities in entropy to exist between initial and final singularities there will be a need for other physics or new explanations, e.g. Weyl Curvature Hypothesis, etc [7]. The justification for such new physics will need to be well founded considering the well established concepts of the earlier physics. Nevertheless a valid way of paraphrasing Penrose's excellent arguments is that a remodeling of the structure and formation mechanism of final space-time singularities will be imperative if time-symmetry is to be preserved and their entropy is to be low or near zero [7]. With current black hole modeling, time symmetry is not possible because their entropy will be high. Fortunately, GR provides us with other routes and mechanisms at arriving at singularity formation without the necessity for infinite contraction of matter to zero radius [9,10,11]. Perhaps such routes need to be exploited in modeling formation mechanisms and structure for black holes, especially since singularities so formed exhibit all the properties of black holes including gravitational attraction without their necessarily having mass. Elaborate discussion on these problems may not be appropriate for this paper. However one problem usually discussed in the literature is worth brief commentary. This is the 'book-keeping' difficulty that arises when discussing the ordinary second law of thermodynamics or even the generalized second law (GSL) [1,3] concerning what happens to the total entropy of the universe when matter (with its own entropy) is dropped into a black hole. The easier solution favored by Bekenstein *et al* is to hold that there is a compensatory increase in black hole entropy thus preventing what will amount to a reduction in universal entropy and a violation of thermodynamic laws [1,3]. However if black hole entropy must remain at zero, i.e. if zero entropy is a characteristic of black holes either because the hole remains "extremal" or because the singularity is not formed of matter (as in [9,10,11]) or for other reasons canvassed in this paper and if it is held that the entropy of the dropped matter is not irretrievably lost so as not to contravene the second law then there is a need for a more imaginative solution. We can therefore conceptualize the hole observationally getting smaller and yet retaining its zero entropy, but making room or space-time available for the existence of the entropy of the matter dropped into it. In this way the second law or GSL is preserved since the entropy of the matter does not then just disappear into oblivion and all entropy can be observationally accounted for. Literally speaking we may contemplate the hole as getting "filled up", when matter-energy falls into it rather than it getting "deeper". There is therefore a need for new conceptual frameworks to understand black hole thermodynamics if only to avoid some of the paradoxes encountered in current approaches.

The central statistical thermodynamic equation for an isolated system with energy E and N bodies or statistical constituents distributed within a number of partitioned compartments or a volume, V is written as

$$S = k \log_e W \qquad (1)$$

where $S$ stands for the entropy, $k$ is Boltzmann's constant, $\log_e$ stands for natural logarithm and $W$ is the number of possible states that a system can assume ($W$, *wahrscheinlichkeiten* = probabilities) and where in addition we can take $W$ as equal to $V^N$. As this is not a detailed discussion on thermodynamics or statistics, only some pertinent points basic to statistics and the evaluation of entropy will be worth mentioning. The first is that only one probable state of the system is possible at a time, i.e. all the possible states cannot occur at the same time. Secondly, the higher the number of statistical constituents, N (e.g. number of balls in V number of partitioned compartments or number of gas molecules in a closed box of volume V), the more the number of different arrangements and possible states, $W$ that can be assumed. Thirdly, it must take the same amount of time to assume each of the available states so all available states remain equally probable. Finally, if it takes an infinite amount of time to assume a different state, only one state will be possible and $W$ will be equal to one. For balls thrown into a partitioned compartment whose clock is tuned to infinite proper-time for example, only one state can be realizable. For more than one state to be realized the balls must have a velocity which the infinite proper-time condition in it has effectively constrained to zero, as it takes an infinite amount of time to transit any given displacement. Using the gas-in-a-box model as another example, if some quantity of gas is placed in a corner of the box, we expect that with time running normally it would spread throughout the box and attain thermal equilibrium. If however, as GR teaches us, clocks can be set to run at different rates and we set the clock within the box to infinite proper-time, we can imagine what consequences should arise. First, in that box, a light beam which hitherto traveled 3 X $10^8$ meters from one end of the box to another in one second, will be infinitely delayed and now take an infinite amount of time to travel the same distance. Secondly, for the gas molecules to spread and be able to reach the other end of the box they will have to travel faster than the speed light has in that box and so the gas molecules remain piled up in the corner of the box as they were placed. Thus when proper-time is set to infinity, it takes an infinite amount of time to assume any other possible states. Adding therefore to what the third law of thermodynamics states, we can say that statistically if there are no alternative states or if there are, but there is no energy available or it takes an infinite time to attain them, then *W* will be one.

The second law of thermodynamics which deals with the concept of entropy gives us an idea of what the entropy of at least one gravitational singularity, the Big bang, could be by stating that in a closed system a lower entropy state precedes a higher one. That being so we may reasonably conclude within thermodynamic laws alone that at least in the case of the Big bang, its entropy is about zero if no lower entropy state precedes it. Mathematically, we can write the second law as

$$\Delta S = S_2 - S_1 = k \log_e W_2/W_1 > 0 \qquad (2)$$

where terms with the subscript 1 represent earlier states and those with subscript 2 represent latter states. Singularity theorems of GR may however be needed in addition to clarify what the entropy of the other space-time singularities could be, although from the time-symmetrical dynamical equations involved, all singularities should have identical entropy, i.e. zero [7,13,14].

It is useful that this property of infinite proper-time is universally present in all gravitational singularities [12,13,14] and so it may be useful to compute the entropy for all types of singularities. From the statistical discussion above, we have already noted that if clocks cannot run within singularities, it will be impossible for the different possible states to be assumed and be manifest in that system. Since, $W$, the number of different states that can be assumed statistically is equal to one then since entropy, $S = k \log_e W$, entropy will be equal to zero. The statistically important logarithmic term is thus present *ab initio* and no ad hoc attempts are necessary to incorporate one.

It could be of importance that entropy be absent in black holes, because if it were present it would increase till equilibrium according to the second law and this could be used as a clock in contravention of GR. And if entropy is present but cannot increase with time to equilibrium so that GR is preserved, the second law will in turn be contravened. Mathematically therefore we can say in black holes

$$\Delta S = 0 \qquad (3)$$

and $W_1$ is always equal to $W_2$. This GR prediction is in conflict with the $\Delta S > 0$ of equation (2). Therefore if GR is correct that clocks stop and if thermodynamic and statistical principles are correct as well that whenever entropy is present it must increase with time till equilibrium, then black hole entropy can only be zero to preserve both GR and thermodynamic principles. An alternative value could contravene either of the two. The third law of thermodynamics also gives some guidance on this. The third law says that as long as the perfect crystal remains at 0K (zero kelvin), $\Delta S$ will equal zero and the crystal will have zero entropy. From this, it is tempting to state that the only values of entropy that preserve the second law of thermodynamics whenever $\Delta S = 0$ are those that are zero, i.e. the only closed systems where entropy does not increase are those where entropy is absent. It can also be noted that when proper-time is infinite as in the gas-in-a-box model, pressure will be zero and therefore temperature too will be absolute zero since there can be no molecular movements in the box. With the third law as a basis we therefore see consistencies that entropy in the box is zero when an infinite proper-time constraint is present. Finally since singularities represent an absence of space-time, either due to being prior to the evolution of collapsed states to uncollapsed ones or being at the end of collapsed states, whatever may eventually be the building blocks or statistical mechanical constituents, N of the system may not be feasibly countable, i.e. N = 0. To summarize the statistical mechanical arguments we can say whenever (i) $\Delta S = 0$ or (ii) T = 0 K or (iii) $W_1 = W_2$ irrespective of the type or source of constraint or (iv) when N = 0 and therefore $W = 1$, entropy, $S$ will likely be zero in value.

There are however views that say that thermal equilibrium has been reached and entropy already maximal in singularities such as black holes, and so thermodynamic laws and GR need not be contravened. Penrose [7] has however already noted some of the paradoxes contained in such alternative ideas. For instance, as he noted, if the Big bang singularity is already in a state of thermal equilibrium and so already maximal entropy, the second law of thermodynamics which says entropy is still on the increase in our universe will be difficult to apply. A careful consideration should show that maximum entropy at thermal equilibrium cannot also be possible in final singularities like black holes without contravening GR. For example at thermal equilibrium there is the typical Maxwell's equation for distribution of velocities [17]. A simplified version of this equation is written

$$v_{mean} = d/t = (8kT/\pi m)^{1/2} \qquad (4)$$

where $v_{mean}$ is the mean particle velocity at thermal equilibrium, d is the mean free path before particle collides with another, t is the time taken to travel the path, $k$ is Boltzmann's constant, T is the temperature at equilibrium and m is the mass of the particle. Substituting infinite proper-time in Maxwell's equation gives us absolute zero as the temperature at thermal equilibrium whenever proper-time is infinite. This state cannot therefore be regarded as one of maximal entropy. Although thermal equilibrium is commonly used to equate the entropy of stationary black holes this conventional notion of thermal equilibrium may therefore be inapplicable under infinite proper-time conditions. A challenge for advocates of thermal equilibrium and maximal entropy in black holes is whether mixing, spreading and heat transfer by conduction, convection or radiation so that equilibrium is reached, can possibly occur in spite of the infinite proper-time present within gravitational singularities, where it takes an infinitely dilated time to cover any displacement. Immediately outside or at the edge of a black hole there may indeed be much disturbance, high temperature and high entropy but what we are concerned about is what is happening at the singularity itself.

The Bekenstein-Hawking formula for black hole entropy [2,3,6] is the most widely used to compute entropy in black holes. Using the formula the preponderance of efforts are made to locate the entropy on the event horizon and not in the deep interior of the hole itself. The extremely peculiar nature of the space and time in gravitational singularity states however seems not to have received sufficient consideration in studying the applicable thermodynamics and this is in spite of the fact that entropy is intricately linked with time evolution. Proper-time values elsewhere in the universe cannot be applicable to calculate the physics in singularities or else we may be faced with which type of clock or value of proper-time to choose from the various gravitational environments present universally. Fortunately, the proper-time applicable in black holes and other gravitational singularities is unequivocally given by GR and the singularity theorems and this is what can be used to study their physics. Using the Bekenstein-Hawking formula, consideration of the peculiar issues involved arrive at this same conclusion of zero entropy for black holes and not non-zero entropy values.

The Bekenstein-Hawking formula is given as

$$S_{bh} = A/4 \times kc^3/G\hbar \qquad (5)$$

where $S_{bh}$ is the entropy, A is the surface area of the event horizon, $k$ is Boltzmann's constant, c is the measured velocity for light, G is the gravitational constant and $\hbar$ is $h/2\pi$, where $h$ is Planck's constant.

Following from equation (5), the Bekenstein-Hawking equation can also be validly written as

$$S_{bh} = Ak/4G\hbar \times (\text{proper distance, s/ proper time, } \tau)^3 \qquad (6)$$

since GR tells us that c = proper distance, s / proper time, $\tau$ \qquad (7)

To make things clearer, from Equation (6)

$$S_{bh} = As^3 k/4G\hbar \times 1/\tau^3 \qquad (8)$$

If GR predicts an infinite curvature at singularities with proper-time, τ going infinite [12,13,14], then calculating using the Bekenstein-Hawking formula as in Equation (8) also gives black hole entropy a zero value.

If it is insisted that we must derive a Hawking temperature for black holes clocks can similarly play a role and in doing this help resolve some of the paradoxes with current approaches like that of a black hole getting colder instead of hotter when a stream of hot radiation flows into it in violation of the zeroth law of thermodynamics. This in itself suggests that the size of a black hole has nothing to do with its temperature. If Hawking's temperature, $T_{bh}$ for a black hole of mass, M is written,

$$T_{bh} = \hbar c^3 / 8\pi k GM \qquad (9)$$

then similarly from Equation (7), $T_{bh}$ is absolute zero when proper-time is infinite.

There is a natural analogue of this correspondence between infinite proper-time and zero entropy which goes to further illustrate the naturalness of the relationship. Using such analogues, especially sonic analogues are recognized, light and sound sharing many things in common and both being propagated forms of energy (e.g. Doppler phenomena, 'dumb holes' [18]). Referring again to the model gas-in-a-box scenario, if the number of gas molecules reduce, the amount of possible states, $W$ reduces as well. At the point when there are no more gas molecules in the box, i.e. vacuum, when N = 0, $W$ becomes one since $W = V^N$. Entropy being $k\log_e W$, is therefore zero as well. In this same state of zero entropy, the time taken for sound to be propagated approaches infinity. The co-existence of zero entropy and infinite time for propagation therefore seems a very natural state of affairs.

In conclusion, although many properties of singularities have been employed to equate their entropy, we see that based on statistical and dynamical principles, their proper-times may seem the most appropriate and accessible to calculate their entropy. Using this property, it is shown first, within statistical principles, secondly when taken into account in the Bekenstein-Hawking formula and thirdly illustrating with a naturally occurring analogue, that the entropy of black holes and all other gravitational singularities is zero as had first been classically speculated. We also propose from a combination of thermodynamic principles and General relativity what may be termed a fourth law of thermodynamics: that the entropy of a system is zero if the proper-time in that system is infinite.

## **Acknowledgements**

I wish to thank Prof. Animalu of the Nigerian Academy of Science along with Dr. Adeloye and Dr. Amaeshi both of the Physics Dept., University of Lagos, Nigeria for stimulating discussions.

# References

1. Wald, R.M., *The Thermodynamics of Black Holes*, Online article, http://www.livingreviews.org/lrr-2001-6 (2001).

2. Bekenstein, J.D., *Phys. Rev.* D, **7**, 2333-2346 (1973).

3. Bekenstein, J.D., *Phys. Rev.* D, **9**, 3292-3300 (1974).

4. Bardeen, J.M., Carter, B. and Hawking, S.W., *Commun. Math. Phys.*, **31**, 161-170 (1973).

5. Carter, B. Black hole equilibrium states, in *Black Holes*, Eds DeWitt, C. and DeWitt, B.S., 56-214, Gordon and Breach, New York, (1973).

6. Hawking, S.W., *Commun. Math. Phys.*, **43**, 199-220 (1975).

7. Penrose, R., *The Emperor's New Mind*, Oxford University Press, New York (1990).

8. Misner, C.W., Thorne, K.S. and Wheeler, J.A., *Gravitation*, Freeman, San Francisco (1973).

9. Khan, K and Penrose, R., *Nature*, **229**, 185 (1971).

10. Matzner, R.A. and Tipler, F.J., *Phys. Rev.* D, **29**, 1575 (1984).

11. Chandrasekhar, S. and Xanthopoulos, B.C., *Proc. Roy. Soc. (London)*, **408**, 175 (1986).

12. Kenyon, I.R., *General Relativity*, Oxford University Press, New York (1980).

**13.** Penrose, R., *Phys. Rev. Lett.*, **14**, 57-9 (1965).

**14.** Hawking, S.W. and Penrose, R., *Proc. Roy. Soc. (London)*, **A314**, 529-48 (1969).

**15.** Cook, A.H., Experiments on gravitation in *300 Years of Gravitation*, Eds. Hawking, S.W. and Israel, W., Cambridge University Press, Cambridge (1987).

**16.** Wald, R.M., Black holes and thermodynamics [gr-qc / 9702022].

**17.** Emiliani, C., *The Scientific Companion*, Wiley, New York (1988).

**18.** Unruh, W.G., *Phys. Rev*. D, **51**, 2827-2838 (1995).